\documentclass[letterpaper,twocolumn,10pt]{article}
\usepackage{usenix,epsfig}
\usepackage{epstopdf}
\usepackage{amsmath}
\usepackage{graphicx}
\usepackage{subfigure}
\usepackage{url}

\usepackage{multirow}
\usepackage{listings}

\usepackage{listings}
\usepackage{array}
\newcolumntype{L}[1]{>{\raggedright\let\newline\\\arraybackslash\hspace{0pt}}m{#1}}
\newcolumntype{C}[1]{>{\centering\let\newline\\\arraybackslash\hspace{0pt}}m{#1}}
\newcolumntype{R}[1]{>{\raggedleft\let\newline\\\arraybackslash\hspace{0pt}}m{#1}}

\begin{document}
\title{Formal Black-Box Analysis of Routing Protocol Implementations}

\author{
 {\rm Adi Sosnovich ~ Orna Grumberg}\\
 \small Computer Science Department \\
 \small Technion -- Israel Institute of Technology
 \and
 {\rm Gabi Nakibly}\\
 \small Rafael -- Advanced Defense Systems Ltd.
} 

\maketitle

\thispagestyle{empty}

\begin{abstract}

The Internet infrastructure relies entirely on open standards for its routing protocols. However, the overwhelming majority of routers on the Internet are proprietary and closed-source. Hence, there is no straightforward way to analyze them. Specifically, one cannot easily and systematically identify deviations of a router's routing functionality from the routing protocol's standard. Such deviations (either deliberate or inadvertent) are particularly important to identify since they present non-standard functionalities which have not been openly and rigorously analyzed by the security community. Therefore, these deviations may degrade the security or resiliency of the network.

A model-based testing procedure is a technique that allows to systematically generate tests based on a model of the system to be tested; thereby finding deviations in the system compared to the model. However, applying such an approach to a complex multi-party routing protocol requires a prohibitively high number of tests to cover the desired functionality. We propose efficient and practical optimizations to the model-based testing procedure that are tailored to the analysis of routing protocols. These optimizations mitigate the scalability issues and allow to devise a formal black-box method to unearth deviations in closed-source routing protocols' implementations. The method relies only on the ability to test the targeted protocol implementation and observe its output. Identification of the deviations is fully automatic. 

We evaluate our method against one of the complex and widely used routing protocols on the Internet -- OSPF. We search for deviations in the OSPF implementation of Cisco. Our evaluation identified numerous significant deviations that can be abused to compromise the security of a network. The deviations were confirmed by Cisco. We further employed our method to analyze the OSPF implementation of the Quagga Routing Suite -- a popular open source routing software. The analysis revealed one significant deviation. Subsequent to the disclosure of the deviations some of them were also identified by IBM, Lenovo and Huawei in their own products.
\end{abstract}

\section{Introduction} \label{sec:introduction}
The Internet owes much of its success to open standards. These standards are being developed in an iterative and open process. They are the fruit of extensive deliberations, trial implementations, and testing. Furthermore, open standards are thoroughly documented and freely available, so they can be readily scrutinized at any time even after their creation. It is generally believed that open standards led to a more robust and secure Internet. Routing protocols are a prime example of open standards. They are a critical part of the Internet infrastructure, allowing seamless interoperability between separate networks.

In stark contrast to the open nature of routing protocol standards, the Internet infrastructure predominantly relies on proprietary and closed-source routers made by large vendors like Cisco. A router's vendor can add, remove or alter the standardized functionality of a routing protocol as it sees fit, as long as interoperability with other vendors' routers is preserved.  Even so, it is not uncommon to have two routers of different vendors that, under some networking scenarios,  cannot co-operate seamlessly~\cite{chen2006can}. Vendors have several possible motivations for deviating from the standardized functionality~\cite{adamczyk2008non}: development cost reduction,  optimization of the protocol functions, or increasing customers' switching cost to other vendors. Additionally, inadvertent deviations may rise due to misunderstanding of the standard or failure to implement it completely.

Identifying these deviations is crucial to assessing their full impact on a network's resilience and security. But the routers' closed source makes this a difficult challenge for the security community. To address this challenge, we leverage formal analysis methods that assist in identifying deviations of a routing protocol implementation from its standard. Our analysis is black-box: access to the implementation's source code or binary code is not required. We only assume the ability to send packets to the router and observe its external behavior. This includes the packets sent by the router and information explicitly available through its user interface. Its black-box nature makes our analysis applicable to any router with minimal changes.

We use a model-based testing approach \cite{apfelbaum1997model,utting2012taxonomy} in which a reference model of a system under test (SUT) is formulated. The model embodies the desired functionality for that system and serves as the basis for test generation. Each test has a desired outcome as determined by the model. The tests are then executed against the SUT and the resulting outcome is compared to the desired result. In our case, the model is formally defined according to the protocol's standard. The SUT is the router's implementation of the protocol, and a failed test indicates a deviation of the implementation from the standard. We use concolic execution \cite{godefroid2005dart,sen2006cute} to automatically generate tests from our model. Concolic testing is a dynamic symbolic execution technique for systematically generating tests along different execution paths of a program. It involves concrete runs of the program over concrete input values alongside symbolic execution. Each concrete execution is on a different path. The paths are explored systematically and automatically until full coverage is achieved.

The model-based testing approach has been successfully employed to find bugs in open source software as well as in open-source implementations of one-to-one network protocols such as TCP and UDP \cite{bishop2005rigorous}. However, routing protocols involve multiple participants. In the realm of these complex multi-party protocols, model-based testing can not be practically applied due to scalability issues. The functionality of routing protocols depends on the dynamics between the participants, their relative locations in the network, and the role each participant plays. A certain protocol may expose parts of its functionality only in specific complex interactions between the participants. Therefore, the number of tests required to verify the protocol's functionality may be prohibitively high.

We propose practical optimizations to the model-based testing procedure that significantly reduce the number of tests generated while still covering the entire functionality of the model. Our main optimization merges different tests that pass through a joint intermediate state. Namely, we merge two long test scenarios that reach the same intermediate state into a single shorter test scenario that starts from the intermediate joint state. This optimization is especially useful for test scenarios in which multiple packets are sent. For example, consider two non-identical sequences of packets, $P1$ and $P2$, that are sent during two test scenarios, $t1$ and $t2$, respectively. Assume that the model ends up in the same final state following each of the two tests. Therefore, a test having a sequence of packets of the form $P1||P$ (the sequence of packets in $P1$ followed by a sequence of packets in $P$) can be merged with a test that has the sequence $P2||P$. The merged test shall have a sequence of packets $P$ and it should be executed from an initial state that is identical to the intermediate joint state of the original two tests.

Our optimized method allowed us to implement the first practical tool to automatically identify deviations in black-box implementations of one of the most complex and widely deployed routing protocols on the Internet -- OSPF (Open Shortest Path First)~\cite{ospf_rfc}. The OSPF protocol is a widely used intra-domain routing protocol deployed in many enterprise and ISP networks. We applied the tool to search for deviations in the OSPF implementation of the largest router vendor in the world -- Cisco. We analyzed three different versions of Cisco's implementation of OSPF in IOS\footnote{Cisco's IOS is a software family that implements all networking and operating system functionality in many of Cisco's routers and switches.} and found 7 significant deviations, most of which compromise the security of the network. Two of them were found in the latest version of IOS. The deviations were acknowledged by Cisco. 

To further demonstrate the generality of our tool, we also employed it to analyze the OSPF implementation of the Quagga Routing Suite~\cite{Quagga} -- the most popular open source routing software. The analysis of Quagga revealed one significant deviation. 

To summarize, our contributions are as follows:
\begin{enumerate}
	\item We propose efficient optimizations to the application of model-based testing to routing protocols.
	\item We devised the first practical tool that automatically identifies deviations in black-box implementations of the OSPF routing protocol.
	\item We found multiple logical vulnerabilities in  widely used OSPF implementations by Cisco, Quagga and others.
\end{enumerate}

The remainder of the paper is organized as follows, Section~\ref{sec:preliminaries} gives background on symbolic execution and concolic testing. Section~\ref{sec:procedures} discusses the formal procedures we use to analyze an implementation of a routing protocol and the optimizations we employ to reduce the number of tests. In Section~\ref{sec:OSPF_procedures} we describe the application of the formal procedures to a tool that analyzes OSPF implementations. In Section~\ref{sec:evaluation} we describe the evaluation of the method against Cisco's and Quagga's OSPF implementation and detail the deviations we discovered. Section~\ref{advatagesnandlimitations} discusses the advantages and limitations of our approach. Finally, Section~\ref{sec:relatedwork} presents related work and Section~\ref{sec:conclusions} concludes the paper. 

\section{Preliminaries} \label{sec:preliminaries}
\subsection{Symbolic execution}
Symbolic execution \cite{cadar2013symbolic} allows analyzing the execution paths of a program and generating corresponding test cases. The input variables of the program are defined as symbolic variables. Then, the program is symbolically run, where symbolic expressions represent values of the program variables. On each execution path a constraint is obtained by collecting all the symbolic expressions that correspond to conditional branches on that path. The path-constraint is a quantifier-free first-order formula over the symbolic variables.
Its solutions form a set  of concrete values of the input variables for which the program runs via the same execution path.
 A test that covers this path is then derived from this solution, containing concrete values of the input variables.

\textit{Concolic testing} (\cite{godefroid2005dart,sen2006cute}) is a dynamic symbolic execution technique to systematically generate  tests along different execution paths of a program. It involves concrete runs of the program over concrete input values alongside symbolic execution. Initially, some random concrete input values are chosen. During a run of the program with this input, symbolic constraints are gathered over the conditional branches of the current execution. Thus, at the end of the run the symbolic path-constraint is obtained. A constraint solver is then used to construct the next concrete execution on a different path. 
This can be achieved, for instance, by negating the last conjunct on the path-constraint not already negated. A new solution for the variant of the path-constraint with negations should necessarily steer a new concrete execution over a different path. This process is repeated systematically and automatically. Finally, the process terminates based on some time limit, coverage criteria, or when full coverage is achieved.



\subsection{Threat Model}
Our analysis is security-motivated, namely we seek to find deviations in routing protocol implementations that may compromise the security of the network. To serve this purpose we design the model, on which we base our testing, such that it incorporates an attacker model that fits a plausible threat to the network. We adopt the common threat model found in the literature (\cite{Wang97, Wu99, Jones06, NDSS12}). This model assumes the attacker has the ability to send routing advertisements to any router within the routing domain. This assumption can be trivially achieved by an attacker that gained control over a single router within the routing domain. The attacker can gain control of a router, for example, by remotely exploiting an implementation vulnerability on the router. Several such vulnerabilities have been published in the past (e.g., CVE-2010-0581, CVE-2010-0580, and CVE-2009-2865).

The attacker's goal is to poison the routing tables of other routers. Because the attacker would like to control the routing domain for an extended period of time, the poisoning should be persistent. Namely, the attack's effects will not be immediately reverted once the attacker finishes executing the attack. 

\section{Black-box analysis procedures} \label{sec:procedures}



In this section we describe the procedures that compose our method of analyzing a routing protocol. We use a model-based testing approach. The model includes a network topology, where each node executes its modeled protocol. We test the protocol's functionalities by systematically generating protocol messages and sending them to nodes within the network topology. Each sent message triggers a different execution path of the model. The details of the method's flow are given in Section~\ref{sec:methodflow}.

Generally speaking, the described method flow can be applied to any network protocol; however, a naive application of this method to a complex multi-party routing protocol will have severe scalability issues in the concolic execution phase, known as the path explosion problem. In this phase a prohibitively high number of tests may be generated. In Section~\ref{multiple} we propose optimizations that reduce the number of tests generated without loss of functionality cover of the model.

\begin{figure}
	\begin{center}
		\includegraphics[width=0.45\textwidth]{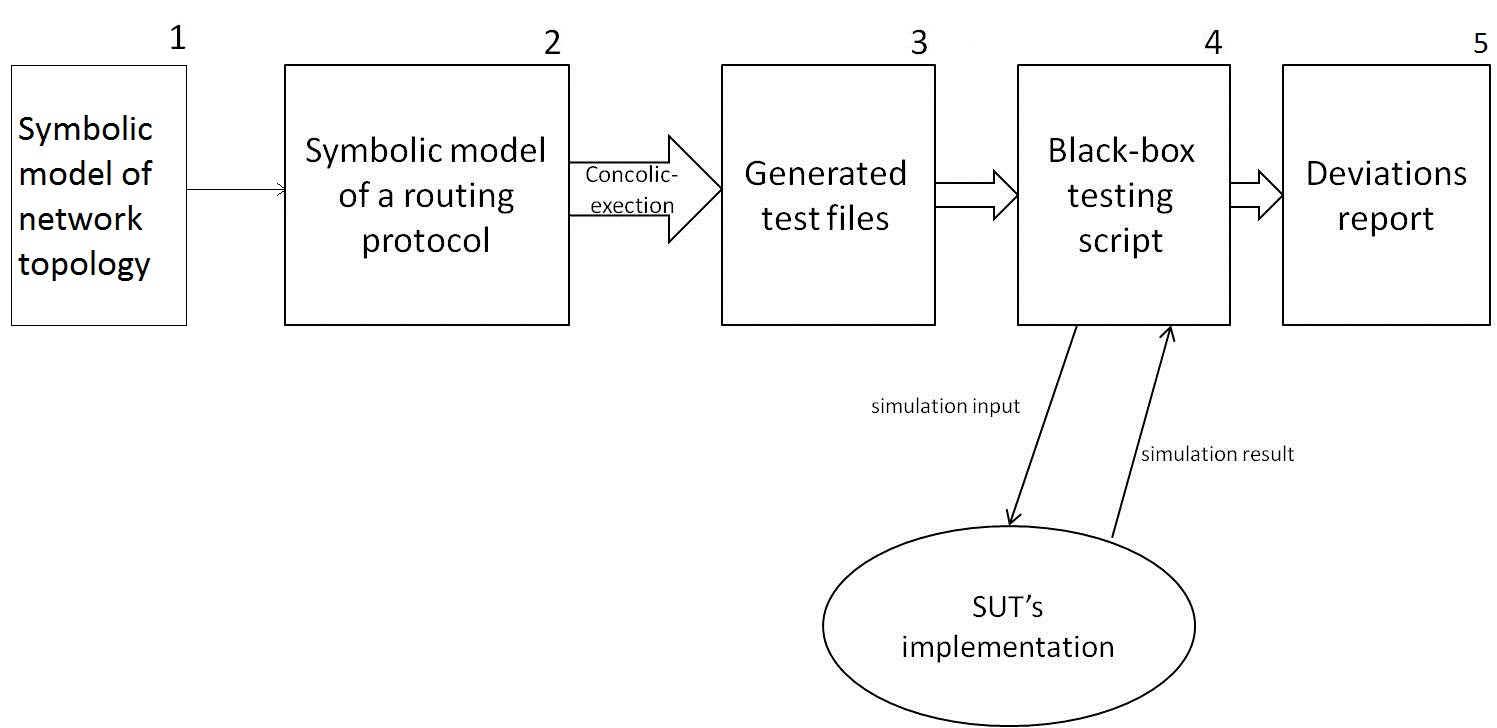}
		\caption{The flow of our method}
		\label{fig:method}
	\end{center}
	
\end{figure}

\subsection{The method flow} \label{sec:methodflow}

The method flow is depicted in Figure~\ref{fig:method}.
Below we explain each numbered box in that figure:
\begin{enumerate}
	\item Produce a symbolic model of a network topology on which the protocol will be executed. Symbolic topology can represent any concrete topology. It allows us to generate tests of the protocol under various topologies, thereby covering specific protocol behaviors that are topology-dependent. The symbolic topology is fed to the next stage in the process. We further elaborate on the symbolic model of the network topology we use in Section~\ref{sec:networktopology}. 
	\item Given a symbolic network topology,  the   model produces a run of the protocol.	
	Each execution path of the model is represented by a concrete run that starts from the \textit{standard initial state } on the chosen topology.
	In the standard initial state, all the nodes are consistent and stable.
	Additionally, all their incoming message queues are empty.
	During the run, messages of the protocol are sent to the nodes, and the run terminates in a stable state after no messages are sent between any nodes on the network.
	
	
	
	
	\item Applying the  concolic execution tool on the  symbolic  model of the routing protocol generates a test file for each execution path of the model. Each test file contains the sent message and the initial and final local states for all nodes in the network.
	Each execution path is finite and reaches a stable state in which there are no more messages that need to be sent between the routers.
	

	\item Each generated test file is executed on the SUT. During the test execution the routers are activated and initialized according to the model's initial state. Then one or more messages are sent to the routers, and after the routers' states become stable again, they are read and compared to the expected state obtained from the model. If the existing state of the routers does not match the model's expected state, the test fails.

	\item   A failed test represents a deviation of the SUT's implementation from the protocol standard. The failed test is accompanied with traces of all messages exchanged between the routers during the run of the test, both on the model and on the SUT. Comparing these traces facilitates the analysis of the deviation.
	
	
\end{enumerate}

The main advantages of the above method is as follows:
\begin{enumerate}
	\item Automatic -- once the protocol's model and topology are determined, the identification of deviations is fully automatic. A deviation may pose a security vulnerability in the SUT, but this is not always the case. Further manual analysis is required to infer what kind of effect the deviation has on security of the implementation.
	\item Full coverage -- the generated tests cover the entire functionality of the protocol's model. This allows to test the model against relatively small topologies with few routers while still being able to check obscure corners of the protocol. Our evaluation in Section~\ref{sec:evaluation} illustrates this very nicely.
	\item Modularity -- the analyzed model need not detail the protocol in its entirety. The model may only include parts of the protocol deemed relevant to the security analysis or parts that may be considered more prone to deviate from the standard. The model may abstract away irrelevant details or even omit them entirely. Furthermore, the analysis may also be split into separate stages, each focusing on a specific part of the protocol.

\end{enumerate}



\subsection{The topology symbolic model} \label{sec:networktopology}
We model two types of network links: point-to-point and multiple access. The former connects only two routers while the latter connects any number of routers. Generally, the two link types are handled differently by routing protocols in terms of the routing advertisements describing them.  We assume the topology has a predetermined number of routers and a predetermined number of multiple access links, denoted by $n$ and $m$, respectively. A symbolic topology model represents any topology that has $n$ routers and $m$ multiple access links. We do not constrain the number of point-to-point links; it can be any number between 0 and $\frac{n(n-1)}{2}$. Two routers can connect via at most one point-to-point link and any number of multiple access links (each multiple access link may connect a different subset of the routers). 

Our aim in implementing the topology symbolic model is to show that our model need not be constrained to specific predetermined topology. Rather, tests can be generated based on execution paths that rely on different topologies, thereby increasing the number of states reached. The downside of using symbolic model for the topology is scalability issues. The symbolic model exponentially multiplies the number of tests compared to a static topology. We discuss this further in Section~\ref{sec:evaluation}.

\subsection{Optimizations}

\label{multiple}

%

In this section we describe our optimizations to the method above. They focus on reducing the number of tests that include multiple messages. It should be noted that running tests on the SUT is relatively fast, even for long test sequences. The heavy part of our method is the generation of tests on the model, due to the path explosion problem. Our optimizations manages to reduce this effort significantly.


The straight-forward approach to generating tests with more than one message would be to use several instances of symbolic messages in the model. However, the disadvantage is that the number of symbolic variables is multiplied in the number of symbolic messages. This may result in the path-explosion problem due to the exponential growth in the number of model paths to execute. Furthermore, this may result in generation of many redundant tests along different execution paths. 

\begin{figure}
	\begin{center}
		\begin{tabular}{ll}
			\hline
			1.	& $ RIS = \{standard\} $ \\
			2.	& $ ERS = \phi $ \\
			3.	& $ k = 1 $ \\
			4.	& while ($ k < K $ )\\
			5.	& \{ \\
			6.	& \quad generated-tests = applyMethod(RIS) \\
			7.	& \quad $ RS = $ extract-reachable-states(generated-tests) \\
			8.	& \quad $ ERS = ERS \cup RIS $ \\
			9.	& \quad $ RIS = RS \setminus ERS $ \\
			10.	& \quad $ k = k + 1 $ \\
			11.	&  \}   \\
			\hline
		\end{tabular}
		\caption{Systematic extension algorithm}
		\label{systematicExtension}
	\end{center}
\end{figure}

\paragraph{Merging paths with same intermediate state.} Let $ P_1 $ be the set  of all model paths with one message, starting from the standard initial state.
Let $ P_2 $ be the set of all model paths with two messages, starting from the same standard initial state.
Let $ P_1(S) $ be the set of all model paths within $ P_1 $ that terminate in the model state $ S $.
We observe that all paths in $ P_2 $ that are extensions of the paths in $ P_1(S) $ with a second message $m$ are equivalent with regard to the functionality they cover in the model. Replacing these paths with only one path that include the message $m$ and starts from initial state $ S $ rather than from the standard one will not reduce the coverage of the paths.
Thus, instead of exploring similar execution paths from each final state of each path in   $ P_1(S) $, we can  apply it only once by using  the model state $ S $ as the initial state, and by applying only one symbolic message from that state. Following this observation we can employ the following procedure to reduce the number of paths in the general case where  the maximum length of the paths is $K$ messages. We call $K$ the maximum message depth. A pseudo-code describing this procedure is given in Figure~\ref{systematicExtension}. 
	$ RIS $ is the set of reachable  states from which exploration via concolic execution has not yet been applied. $ ERS $ is the set of explored reachable states from which the exploration of concolic execution  has already been applied.	For each reachable state, we also keep a sequence of corresponding messages from which the reachable state is obtained. For example, consider the pair $ (S,(M_1,M_2)) $, where $ S $ is a reachable state and $ (M_1,M_2) $ are the corresponding messages. The notation means that if a run of the model starts from the standard initial state and these messages are sent one by one, eventually the final state observed on the model is $ S $.
	
	In line 1 of  the pseudo-code we initialize the set of reachable initial states with the standard initial state. In line 2 we initialize the set of explored reachable states with the empty set. $ K $ represents the current depth of the generated tests. It is initialized to 1, since on the first iteration we generate tests with a single message sent from the standard initial state. 
	In each loop iteration for a new depth, lines 6-10 are applied. In line 6 we apply our method from the previous section on every  state in $ RIS $ with a single symbolic message. The generated tests are kept and sent to the black box testing script. To initialize the SUT according to the new initial state, we adjust the initialization process in the black-box testing script. The adjustment requires sending the corresponding sequence of messages that are kept with the reachable state, as part of the initialization.
	Afterward, in line 7, we analyze the generated tests from the previous stage. For each generated test with $ K $ messages we detect its final state. The set of all reachable states (and their corresponding message sequences) from the generated tests are kept in $ RS $. In line 8, we add to the set of explored reachable states the set of the reachable  states from which the method was applied on the current iteration. Then, in line 9, we update the set of reachable  states from which exploration of depth $ K+1 $ should be applied by removing the set of all explored reachable states from the set of the detected reachable states of the last iteration. Finally, $ K $ is increased by 1 in preparation for the next iteration.

	We note that each reachable state may actually have multiple sequences of messages leading to it in the model. We choose to use one representative message sequence out of all possible sequences.
	
	\vspace{-1pt}
	
\paragraph{Arbitrary prefix paths.} We describe now a different optimization that does not guarantee the model coverage. We believe it has merit in analysis scenarios which do not mandate the full coverage of the model. The idea is to explore the model starting from states reachable by an arbitrary number of messages. Such an optimization allows to explore the model in a greater message depth. A reachable state that may require an arbitrary number of messages is applied on the routers, after which a concolic execution with a single symbolic message is applied.	The arbitrary reachable initial state may be achieved by random simulation of the protocol with an arbitrary number of messages, starting from the standard initial state. 
	This process requires adjustment of the initialization process in the black-box script as well. The sequence of messages leading to the new reachable initial state should be provided. 
	Let $ (S,(M_1,M_2,...,M_k)) $ be the reachable state and its corresponding message sequence.
	Let $ {F_1,F_2,...,F_n} $ be the set of test files generated from concolic execution with $ S $ assigned as the initial model state and with a single symbolic message. When running the generated tests on the SUT, we adjust the initialization process by sending the sequence of initializing messages $ (M_1,M_2,...,M_k) $. Then, we  compare the state of the routers on the SUT with the given model state $ S $ and make sure that they match. Only then we can send the message from the generated test file $ F_i $ and compare the final states of the SUT and the test file.
	
It should also be noted that execution time of a test on the SUT is relatively fast, even for long test sequences. Therefore, shortening a test sequence is not useful, but rather to reduce the number of tests as done by the above optimization.

\section{Black-box analysis of OSPF} \label{sec:OSPF_procedures}
Towards the evaluation our method we apply it to a tool that tests one of the most complex routing protocols on the Internet -- OSPF. OSPF is the routing protocol of choice for many ISP and enterprise networks. We focus on its most widely used version -- version 2~\cite{ospf_rfc}.
Before we detail the model we used, we give a brief overview of the protocol.

\subsection{OSPF Background} \label{sec:ospfbackground}
OSPF (Open Shortest Path First) is an intra-domain routing protocol,  used within collections of networks, each of which is called an autonomous system (AS). OSPF is a link state routing protocol:  each router advertises a message called a Link State Advertisement (LSA), containing its links to neighboring networks and routers and their associated costs. Each LSA is flooded throughout the AS. Routers construct a complete view of the AS topology by compiling all the LSAs they receive into a single database (LSDB). From this global view routers compute their routing tables. Each router is identified by the IP address of one of its interfaces, called a router ID.

A local network having exactly two routers directly attached to it is called a point-to-point link. Each of the two routers advertises a link to its peer. In contrast, a local network having two or more routers is called a transit network. A router connected
to a transit network advertises a link to the network
rather than to the neighboring routers. In addition,
one of the neighboring routers is chosen to act as a designated
router. This router advertises an LSA on behalf
of the local network, in addition to its own LSA, advertising
links back from the network to all the routers
attached to the network (including itself).
An LSA describing the links of a router is called a Router-LSA. An LSA describing the links of a transit is called a Network-LSA.

\begin{figure}%
	\centering
	\includegraphics[width=1\columnwidth]{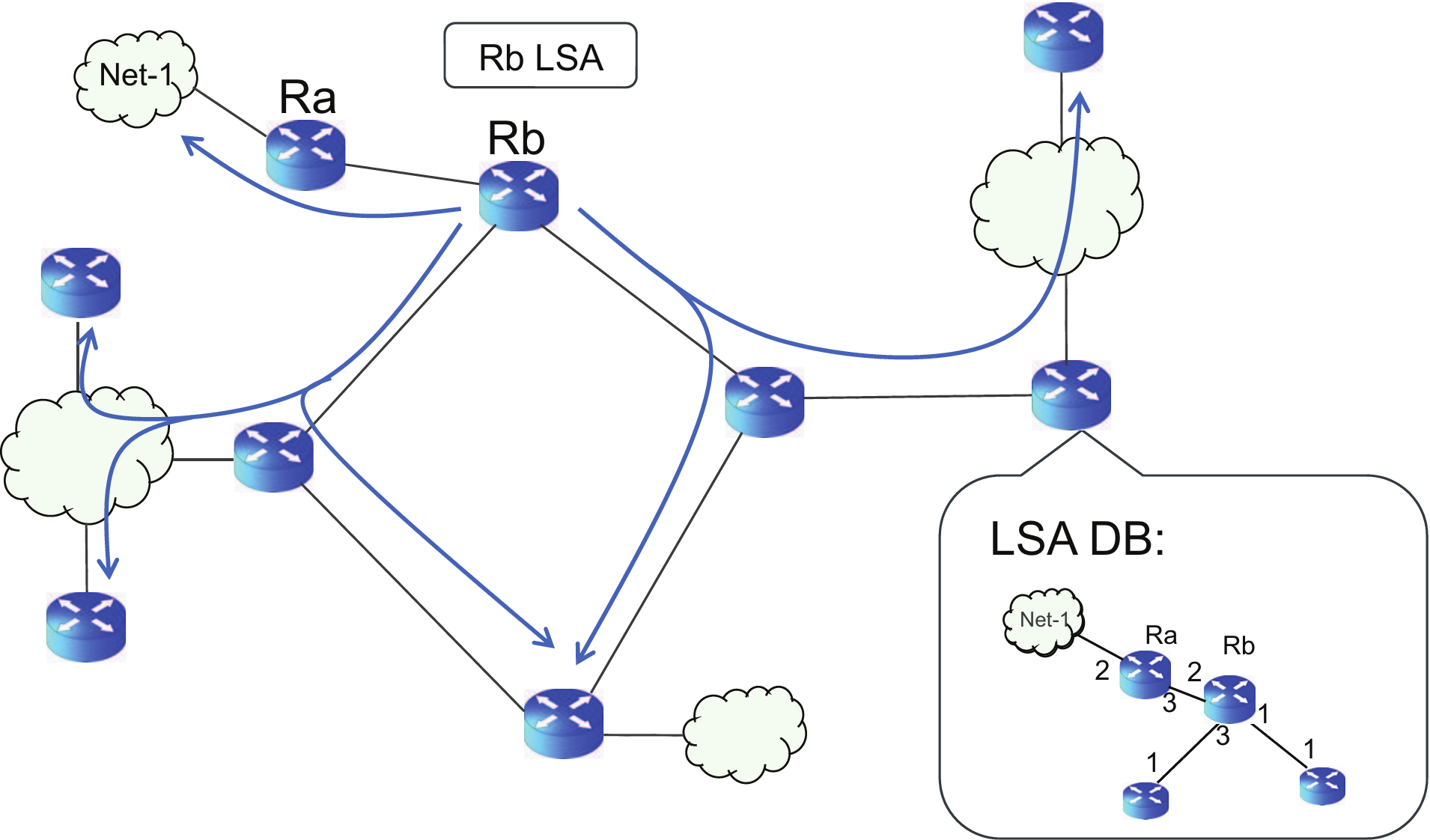}%
	\caption{An example of LSA flooding (taken from \cite{NDSS12})}%
	\label{simpleospf}%
\end{figure}

Figure~\ref{simpleospf} illustrates the flooding of an LSA throughout the AS while the routers build their LSA database (LSDB) to construct their view of the AS topology.

OSPF messages are sent directly over IP. Therefore, OSPF must employ its own mechanisms to ensure reliable transmission of messages. Once an LSA is received from a neighboring router, an acknowledgment is sent to that router. Once a router detects a change in its local topology (e.g., it has a new link or one of its links has gone down), it sends out a new instance of its own LSA with the new information. Moreover, a new instance of an LSA is advertised periodically every 30 minutes, by default, even if there is no change. An LSA includes a LS sequence number field, which is incremented for every new instance. A fresh LSA instance with a higher sequence number will always take precedence over an older instance with a lower sequence number. In addition, an LSA includes an Age field indicating the elapsed time since the LSA's origination. When it reaches 1 hour, the LSA instance is removed from the LSDB.

The sequence number field is 32 bits long. Once the sequence number reaches its maximum value, it needs to wrap to zero in the next LSA instance. To do that the LSA instance with the maximum sequence number (MaxSeqNum) is first flushed by advertising another instance having the maximum sequence number and an age field of 1 hour. This instance replaces the current LSA instance but is then immediately removed from the LSDB due to its age. Therefore, no instance of that LSA is kept in the LSDB. At this stage a fresh instance of the LSA with an initial sequence number will be advertised.

One of the dominant defense mechanisms of OSPF against rogue routers residing within the network is the \emph{fight-back} mechanism. Once a router receives a false instance of its own LSA, it immediately advertises a newer instance which cancels out the false one. The newer instance will have a sequence number that is incremented by one as compared to the sequence number of the false instance. This should prevent an attacker from persistently and stealthily falsifying an LSA of a router the attacker does not control. Since each LSA is flooded throughout the AS, we are assured that in the general case the correct fight-back instance will be received by all routers in the AS.

An LSA header is composed of the following fields:
\begin{itemize}
	\item LS age -- The time in seconds since the LSA was originated.
	\item LS type -- The type of LSA (e.g., Router, Network)
	\item Link State ID -- Identifies the part of the AS that is being described by the LSA.
	\item Advertising Router -- Identifies the router that originated the LSA.
	\item LS sequence number -- The sequence number of the LSA.
\end{itemize}

\subsection{The OSPF symbolic model}
\label{model}
Our black-box analysis is security-motivated. Hence, our aim is finding in an OSPF implementation deviations from the protocol's standard that may give rise to security vulnerabilities. We focus on the most important type of attack against OSPF: a rogue router advertising false LSAs on behalf of other routers in the routing domain. Such an attack -- if executed successfully -- allows the attacker to significantly poison the routing tables of all routers in the routing domain. This allows an attacker to reroute packets, create routing loops, and cut off connectivity in the network.

To serve the purpose of finding security vulnerabilities our OSPF model focuses on the core parts of the protocol that are relevant to its security against the above type of attack. These parts include the LSA flooding procedure, the fight-back mechanism, the LSA message structure, and the LSA purge procedure. We leverage and extend an OSPF model that was proposed by \cite{DBLP:conf/cav/SosnovichGN13} and \cite{nakibly2014ospf}. The model was previously used in the context of model checking to find vulnerabilities in the protocol's standard itself. In the following, we will refer to our model as the OSPF \textit{reference model}.


Below we detail the main aspects of the OSPF reference model that we implemented and used for the black-box analysis.

\paragraph{The modeled LSA structure}
An LSA in our model contains the following fields:
LS type, Link State ID (LSID), Advertising Router (AR), Sequence Number (SeqNum), LS age, and Links List. The LS age is abstracted into a Boolean flag indicating whether it is MaxAge or not. An LSA message contains the LSA itself, and in addition the IP packet's source and destination.

\paragraph{State} A state in our model is the set of  LSDBs of all routers and the state of the routers' incoming queues.
A state is considered stable when all routers' queues are empty.

\paragraph{The symbolic variables}

We use symbolic variables within the LSAs that are to be sent by the attacker, within the LSDBs of the routers' initial states, and to define the network topology in which the routers and links are arranged. Below we detail the symbolic variables and their domains.

\begin{itemize}
	
\item \textbf{LSA:} Each sent LSA message  has the following symbolic variables:

\begin{itemize}
	\item Sequence number -- the seqNum field is a number in the range $ [0,maxSeq] $, where maxSeq is a predefined constant in the model.
	\item Destination -- the destination field is one of the routers within the chosen topology.
	\item Advertising Router -- the AR field is the router ID of any router within the network topology.
	\item LSID -- the LSID field is the router ID of any router within the network topology.

\end{itemize}

The symbolic LSA message is of the form:
\[  \langle type,src,dest,LSID,AR,seq, links, age  \rangle  \]

The remaining fields, which are not symbolic, are concretely assigned as follows:
\begin{itemize}
	\item $ type=routerLSA $
	\item $ links=[]$, i.e.,  an empty list of links
	\item $ src $ corresponds to $ dest $:  The $ dest $ field determines the $ src $ field, as we know on the given topology through which neighbor the LSA is routed to is destination.
	
	\item $ age = 0 $  (assigned as not MaxAge).
	
\end{itemize}

\item \textbf{LSADB:} The LSDBs are initialized in the standard initial state.
The sequence numbers of the LSAs are symbolically initialized with additional symbolic variables. Their range is $ [0,k] $, where $ k $ a predefined constant smaller than MaxSeqNum.
\item \textbf{Topology:} The existence of a point-to-point link between each pair of routers is defined by a Boolean symbolic variable. We have $\frac{n(n-1)}{2}$ such variables (where $n$ is the number of routers). The attachment of a router to a multiple access link is defined by a Boolean symbolic variable. We have $nm$ such variables (where $m$ is the number of multiple access links).

\end{itemize}

\paragraph{The main function}
The main function of the model has input symbolic arguments as described above.
The run of the model starts with initialization of the LSDBs according to the standard initial state.
Afterwards, for each symbolic LSA, the LSA is sent to its destination and then a loop is applied. On each loop iteration every router runs its procedure once.

\paragraph{The router procedure}
When a router $R$ receives an LSA, it checks whether the  LSA exists in its LSDB. If it does not exist or is considered newer than the existing instance, the router floods the LSA and updates its LSDB accordingly. If the LSA is self-originated,  a fight-back is triggered. If the sequenceNumber of the fight-back LSA reaches the MaxSeqNum, the router originates an LSA with MaxSeqNum and MaxAge, and then a new  LSA with InitialSeqNum. The MaxAge LSA triggers the other routers to purge $ R $'s LSA from their LSDBs.

\paragraph{The attacker procedure}
The attacking router denotes a compromised router within the topology. We predetermine that the compromised router to be router number 0. No generality is lost due to this definition since our model considers a symbolic topology. During the execution of the model the attacker sends the sequence of symbolic LSAs according to their predefined order.

\paragraph{Interleaving}
The model simulates each router's run sequentially, in a round-robin scheduling. When a single LSA is sent, the interleaving does not affect the final state. As long as only one 'external' LSA is being sent among the routers, all interleavings of subsequent LSAs result in the same final state.
For multiple LSAs in the model, we consider  interleavings in which every LSA is sent separately in a specific order. After each LSA is sent, the routers are activated until stabilization is achieved. Only then is the next LSA  sent, and so on.
Thus, for every number of sent LSAs, we can expect that an actual run on the SUT would terminate in a similar final state as in our model.

\paragraph{Test generation}
We use concolic testing procedure\footnote{Our concolic executions are based on a tool called mini-mc~\cite{mini_mc} which we adapted to our use case.} to generate test cases that cover our model's execution paths. We use z3~\cite{z3} as the constraint solver to perform the concolic executions.  The structure of a generated test includes the network topology, the initial state of the routers, the content of the sent LSAs, and the final states of the routers.

\subsection{Model implementation}
We implemented the above OSPF model using Python. The model code consists of roughly 1000 lines of code. The amount of time devoted for the protocol modeling and its implementation was about a couple of weeks. Code 1 gives a high level pseudo-code of the main function in the model and the two principal functions of a router.

\begin{minipage}[t]{0.5\textwidth}
	\centering
\begin{lstlisting}[language=python,firstline=1, lastline=42, basicstyle=\tiny,caption=An overview of part of the model implementation,label={lst:code1}]
def runModel (symbolic_vars):
   init_topology()
   init_routers()
   init_messages(symbolic_vars)
   
   for m in Messages:
		send m by attacker to its dest
		while not stable state:
			for R in Routers:
				R.run_procedure()
	
   gen_output(init_state, messages, final_state);

class Router:
	def run_procedure(self,topology):
		if len(self.queue) >0 :
			m = self.queue.pop(0)
		
			if m.dest != self.ID :           
				nextHopID = self.routingTable.getNextHopID(self.ID, m.dest)
				m.interface = topology.getLinkInterface(nextHopID,self.ID)
				topology.routers[nextHopID].queue.append(m)
			else:                              
				self.handelLSAMsg(m,topology)
		
	def handelLSAMsg(self,m,topology):
		should_flush = False
		if m.age == Age.max and m.AR != self.ID:
			should_flush = True

		if lsa not in self.LSADB:
			self.DB.append(m)  
			self.flood(m,topology)
			return
		if lsa in self.LSADB and is newer instance AND m.AR != self.ID:
			self.flood(m,topology)
			delete old LSA in DB
			self.DB.append(m)
			return
		if m.AR == self.ID:
			self.FightBack()
			return
\end{lstlisting}
	
\end{minipage}

\subsection{Black-box testing of the generated tests}
\label{script}

The black-box testing script interacts with the SUT's OSPF implementation. The script's input is the set of test files to run that was generated in the previous stage. For each test file the script applies a corresponding test run on the SUT. The results of the run are compared with the expected results that are specified in the test file. After completing a run of a test file on the SUT and before applying a new run of another test file, the routers are restarted to avoid side effects from previous runs on the following runs.
 The output of the script is the list of failed tests for which the results of the SUT's run did not match the expected results obtained from the model.

Let $ (mInitialState, mLSAs, mFinalState) $ be the specified initial state, sent LSAs, and final state within a test file obtained from the model.
We detail below the significant procedures applied in this stage by the black-box testing script to allow the comparison between the runs of the OSPF model and the SUT:

\paragraph{Initialization of the routers}

The initial state of the routers on the SUT when they are initially activated or after being restarted is the \textit{standard initial state}, in which the LSDBs of all routers are complete and consistent, containing all of the LSAs originated by the routers in the network topology. The LSDBs correctly reflect the network topology view when they are consistent, and the routing tables of the routers are computed based on the LSDBs content. The sequence numbers of the LSAs are arbitrary in the standard initial state of the routers within the SUT.

Since $ mInitialState $ is the standard initial state, the contents of the LSAs and routing tables of the routers within the SUT after restart should match the model initial state as specified in the test file. However, the sequence numbers of LSAs within $ mInitialState $ have concrete values in terms of the model domains, whereas the LSAs of the routers on the SUT have arbitrary sequence numbers. We have to make sure that the sequence numbers of the routers' LSAs are consistent with  those specified in $ mInitialState $.
This is because the initial sequence numbers of the LSAs may have an effect on the final state.
Let $ mSeq_A, mSeq_B $ be  the initial sequence numbers of two routers $ A,B $ based on a test generated from the model, and let $  seqA, seqB $ be the  initial sequence numbers of the two routers on the SUT.
 It is expected that on the SUT the initial state would match as follows: $ seqA - seqB    =  mSeq_A - mSeq_B $.

A general OSPF implementation does not allow  the sequence numbers of the routers' LSA to be manually set. Therefore, we artificially apply such an initialization by sending each router a self-originated LSA of its own.
Assume, for instance, that routers A and B have the following initial sequence numbers in  $ mInitialState $: $ mSeq_A=2, mSeq_B=0 $, and the arbitrary initial state of the routers' LSAs in the SUT contains the following sequence numbers: $ Seq_A=0x80000005,Seq_B=0x8000000F $. Then, we send to $ A $ an LSA with $seq=0x80000012 $, and to $ B $ an LSA with $seq=0x80000010 $. Thus,  after the fight-back, we finally have a state with  $ Seq_A=0x80000013,Seq_B=0x80000011 $, which matches the initial state of the model.


\paragraph{Sending the symbolic LSA}
To send the LSA from the test file, we take into account the concrete values of the symbolic variables within the generated tests. We use the Scapy framework~\cite{scapy} to generate a corresponding LSA packet to be sent on the SUT. The  sequence number of the  LSA to be sent is based on the initial sequence number of the initial matching LSA on the SUT and on the  sequence number of the sent LSA as specified in the test file. For example, consider a case where $ mLSAs $ contains a sent LSA with the fields:$ AR=1,LSID=1,seqNum=3 $. This means that on the test from the model, an LSA was sent on behalf of $ R_1 $ with $ seqNum=3 $. In order to translate it to an LSA to be sent on the SUT, we need to consider the following values: let  $ mSeq_1=1 $ be the initial sequence number of $ R1 $'s LSA on $ mInitialState $, and let  $ Seq_1= 0x80000005 $ be the initial sequence number of $ R1$'s LSA on the SUT after applying the initialization process in the previous stage. Then, the sequence number of the LSA to be sent on behalf of $ R1 $ on the SUT would be $ Seq_1 + (seqNum - mSeq_1) $, or in that specific case: $0x80000005 + (3-1)  = 0x80000007 $.

\paragraph{Comparison of matching states}
To compare the final LSDBs from the model and from the test run on the SUT, we check that for each LSA in the model LSDB there is a matching LSA in the SUT's LSDB and vice versa. LSAs are considered matching if the following fields  match: LS type, LSID, AR, Links. The links list is considered matching if for every link in the SUT there is a matching link in the model and vice versa.
The sequence numbers specified in the final state of the generated  test file are given as symbolic expressions in terms of the symbolic input variables. Thus, if the test file states that the final expected sequence of the LSA of $ R_1 $ is $ symbR1Initial + 1 $, then we check that the matching LSA from the SUT's concrete run has a sequence number that is larger by 1 than the initial sequence number of $ R1 $'s LSA at the initialization process.


\section{Evaluation} \label{sec:evaluation}
In this section we describe an evaluation of our method against Cisco's IOS implementation of the OSPF protocol. Nonetheless, our method can be similarly and quite easily applied to other OSPF implementations as well. To show this we also tested the Quagga Routing Suite~\cite{Quagga}\footnote{Even though Quagga is open-source we applied the tests in a black-box manner as in the case of Cisco.}. See Section~\ref{sec:results2}.

We have found 7 deviations in Cisco's OSPF implementation and one deviation in Quagga. All deviations were confirmed by the vendors. Although the deviations we found are of interest in and of themselves, our main aim in this section is to verify that the method we propose is indeed efficient and practical for finding protocol deviations even in complex routing protocols standards.

\subsection{Testbed} \label{sec:testbed}
To test Cisco's OSPF implementation we used alternately two network simulation software: GNS3~\cite{gns3} and VIRL~\cite{virl}. Both software suites allow to simulate a network of multiple routers, each running an emulation of an actual IOS image (identical to the images used in real Cisco routers). 

\begin{figure}
	\begin{center}
		\includegraphics[width=0.3\textwidth]{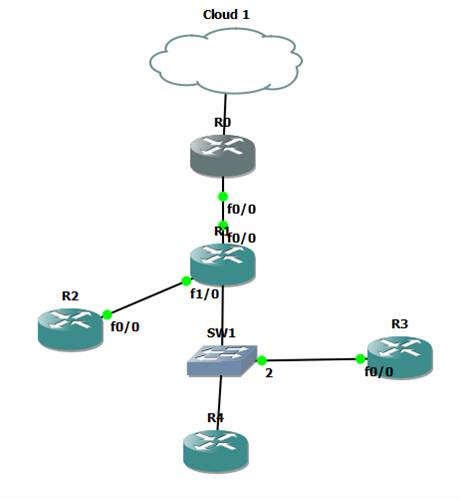}
		\caption{One of the static network topologies against which we ran the evaluation}
		\label{fig:Topology_new}
	\end{center}
\end{figure}

As noted in Section~\ref{sec:networktopology} the symbolic topology model causes the test generation phase to suffer scalability issues. In table~\ref{tbl:symtopo}  we summarize the number of generated tests for small-size symbolic topologies. It is obvious that as the symbolic topology grows the number of generated tests grows exponentially. Very quickly the number of tests even for moderate size networks becomes too large. We shall address the scalability issues due to symbolic topology in future work. For this evaluation we chose to use a handful of static topologies.  One of the main topologies we analyzed is depicted in Figure~\ref{fig:Topology_new}.  We have chosen a simple topology that contains only 5 routers and 5 links. This is with the explicit intention of showing the power of our method with regard to functionality coverage. The full coverage of the model makes it possible to unearth protocol deviations even in simple topologies that may seemingly do not expose the full complexity of the protocol. The attack messages are sent through router R0 (see Figure~\ref{fig:Topology_new}).  

\begin{table}
	\centering
		\begin{tabular}{|c|c|c|}
			\hline
			\# routers & \# multiple access networks & \# tests\\
			\hline
			2 & 1 & 259 \\ \hline
			2 & 2 & 1039 \\ \hline
			3 & 1 & 15029 \\ \hline			
		\end{tabular}
		\caption{The exponential increase in number of tests as the symbolic topology grows}
		\label{tbl:symtopo}
\end{table}

We extract the contents of the LSDB and the routing table from each router by connecting to it using a Telnet or SSH session and issuing the appropriate CLI commands. Every test is preceded by a soft reset of all routers\footnote{To reset a Cisco router the \ttfamily 'reload' \normalfont CLI command is issued.}.

We emulated the routers in those topologies using images of three stable IOS versions as detailed in Table~\ref{tbl:IOSver}. These versions were evaluated due to the large time gap between their release dates -- 5 years in total. This time gap leads us to assume that the there are non-negligible changes in the code base between the three versions, even though  the core functionality of the OSPF standard remained the same during this time period. The changes may be due to new proprietary features, optimizations of protocol functions or bug fixes. These changes in the code base allowed us to verify that our method indeed is capable of identifying different deviations in the different versions of the same vendor's implementation. 

Examining past versions also allowed us to get a sense of the extent of false negatives (missed deviations) of our analysis. We know we missed a deviation if we do not discover a vulnerability that Cisco already announced. Since Mar. 2011 (the release date of the earliest version we tested -- 15.1(4)M) Cisco announced 2 vulnerabilities related to OSPF implementation in IOS (CVE-2013-0149 and CVE-2013-5527)\footnote{Based on a search in NIST's National Vulnerability Database}. Only one of them is related to a deviating functionality with respect to the standard (the other vulnerability was related to a parsing bug). Our tests discovered that vulnerability.

\begin{table}
	\centering
		\begin{tabular}{|c|c|}
			\hline
			IOS Version & Release date\\
			\hline
			15.1(4)M, release software (fc1) & Mar. 2011 \\
			\hline
			15.2(4)S7, release software (fc4) & Apr. 2015 \\
			\hline
			15.6(2)T, release software (fc4) & Mar. 2016 \\
			\hline			
		\end{tabular}
		\caption{Cisco's IOS versions tested for deviations}
		\label{tbl:IOSver}
\end{table}


\subsection{Tests results overview} \label{sec:results}

We illustrate our findings using the topology depicted in Figure~\ref{fig:Topology_new}. We first discuss the performance gain afforded by the optimizations we presented in Section~\ref{multiple}. Then we summarize the deviations we uncovered throughout our analysis. 


\begin{table}
\centering
\begin{tabular}{|C{1.5cm}|C{2.5cm}|C{2.5cm}|}
	\hline  Number of LSAs & \begin{tabular}[c]{@{}c@{}}No optimization\\(\# tests, time)\end{tabular}   & \begin{tabular}[c]{@{}c@{}}With optimization\\(\# tests, time)\end{tabular}  \\
	\hline   1 & 395, 20 min & 395, 20 min \\
	\hline   2 & N/A, $>$24 h & 40188, 2h \\ 	
	\hline
\end{tabular}
\caption{A summary of the number of generated tests with and without the \emph{path merging optimization}}
\label{tbl:results}
\end{table}

\begin{table}
\centering
\begin{tabular}{|C{3cm}|c|C{2.5cm}|}
	\hline   & Time  & \begin{tabular}[c]{@{}c@{}}\# of unique model \\states reached\end{tabular}  \\
	\hline   No optimization & 80 & 2h \\
	\hline   With optimization & 120 & 10 min \\ 	
	\hline
\end{tabular}
\caption{A summary of the time it took to generate 1000 tests and the number of unique model states they cover with and without the \emph{arbitrary prefix paths optimization}. Path prefix length=3}
\label{tbl:results-arbprefix}
\end{table}

Table~\ref{tbl:results} summarizes the number of generated tests and the time it took to generate them while using either 1 or 2 symbolic LSA messages. For both cases we generate the tests with and without the path merging optimization described in Section~\ref{multiple}. The number of generated tests for a single symbolic message is 395. Since the described optimization does not address single LSA scenarios the number of tests remain the same regardless of the optimization. Nonetheless, the necessity of the optimization becomes quickly apparent while generating tests with 2 symbolic messages. We initially tried to generate tests by directly using two symbolic LSAs (without the use of the optimization). However, this resulted in the generation of excessive number of tests. In fact, the concolic execution process took over 24 hours (after this time we stopped its execution). This fully exemplifies the well-known path-explosion problem. The application of a sequence of two symbolic LSAs to the OSPF model resulted in an overwhelming number of execution paths. 

After applying the path merging optimization, we analyzed the reachable states of the generated tests from the previous stage (with a single LSA). We found that out of 395 test files only 103 of them have unique final state. Then, we were able to generate tests from each new reachable state out of the discovered 103 states. In total 40,188 tests were generated for two LSAs using the optimization just under 2 hours.

Table~\ref{tbl:results-arbprefix} examines the performance of the arbitrary prefix paths optimization described in Section~\ref{multiple}. Since this optimization does not guarantee full model coverage we benchmark its benefit differently. Given the same number of allowed tests, we measure the added number of model states reached during the concolic testing phase as compared to a standard application of model-based testing. In this evaluation we assume we have a budget of 1000 tests that we allow to run against the SUT. Without using the optimization this number of tests were generated in 2 hours and reached 80 unique model states. In contrast, when we employed the arbitrary prefix paths optimization using the same budget of 1000 tests we are able to cover 50\% more unique model states. This is due to the fact that the arbitrarty prefix allow us to venture deeper in to the model and reach new states we could not have reach with shorter paths. 

Moreover, an even bigger benefit of the arbitrary prefix paths optimization is the fact that the tests were generate in only 10 minutes (more than 90\% improvement). The time improvement is due to the fact that the concolic executions were run with only one symbolic message (which were prefixed by arbitrary messages). Without the optimization we need to run concolic executions with at least two symbolic messages which is a much more time consuming task. 

In total, throughout our evaluation we found 7 deviations in version 15.1. We consider 6 of them as security vulnerabilities,
3 of which were also reproduced in versions 15.2 and 15.6, while the other 4 deviations were already fixed by Cisco in those versions. Table~\ref{gaps_category} summarizes the number of found deviations per model version and categorizes the deviations according to their impact. 

\begin{table}[]
	\centering
	\begin{tabular}{|l|l|l|}
		\hline
		\textbf{Deviation category} & \textbf{15.1}   & \textbf{15.2 , 15.6} \\ \hline
		Harmed Routing         & \{1,2,3,4,5,6\} & \{1,2\}        \\ \hline
		Affected Stability      & \{2,5,6\}       & \{2\}          \\ \hline
		Non-vulnerability     & \{7\}           & \{7\}         \\ \hline \hline
		\emph{Total \# of found deviations} & \emph{7} & \emph{3} \\ \hline
	\end{tabular}
	\caption{Summary of found deviations in the three IOS versions, categorized by types}
	\label{gaps_category}
\end{table}

In the next subsection we give details of the found deviations.

\subsection{Analysis of found deviations}
\subsubsection{Unknown vulnerabilities}
\label{unknown}
Deviations 1 and 2 from Table~\ref{gaps_category} represent new, previously unknown security vulnerabilities that our analysis revealed in all three Cisco IOS versions that we tested. Cisco acknowledged these vulnerabilities.

\begin{enumerate}
	\item \textbf{Incomplete fight-back for Rogue LSA with maximum sequence number:}
	
		\textbf{Description}:	 A false LSA having the maximum sequence number was sent by unicast to a router $ R $ on behalf of $ R $ itself.
		The router originated an LSA with MaxSeqNum and MaxAge. However, it
		unexpectedly did not originate its LSA with InitialSeqNum. The routing tables of the other routers were affected due to that missing LSA of $ R $.
		
		\textbf{Impact:}  This deviation allows an attacker to send a spoofed LSA that persistently disrupts the routing in the network. The missing LSA origination results in loss of connectivity.

		\textbf{Comments:} This deviation is topology-dependent. It was observed on routers $ R0,R1,R2 $ only. It was not observed at all in a smaller topology that Cisco tested while trying to independently reproduce the deviation.
	
	\item \textbf{Incorrect MaxAge LSA origination during fight-back:}

		\textbf{Description:}  A false LSA having the maximum sequence number was sent on behalf of $ R2 $ to $ R1 $. Consequently, $ R1 $ stored the false LSA in its LSDB and flooded it to $ R2 $, as expected. Since $ R2 $ received a self-originated LSA with MaxSeqNum, it should first originate a MaxAge LSA, and then an InitialSeqNum LSA.
		Based on the standard, the MaxAge LSA should carry the same links as  the MaxSeqNum LSA.
		However, in the tested implementations, $ R2 $ originated a MaxAge LSA with its own valid links, instead of the invalid links from the false LSA.
		After $ R2 $ sent this wrong MaxAge LSA to $ R1 $, $ R1 $ had to check whether the MaxAge LSA is newer than the false LSA in its LSDB. Both LSAs have the same sequence number, MaxSeq.
		Based on the standard, the checksum field is compared in such a case. The false LSA in $ R1 $'s LSDB had a larger checksum value in our test. Thus, this false LSA was considered newer.
		Note that if the MaxAge LSA would have been correct, the checksum values would have been identical, and the MaxAge LSA would have been considered newer, as expected.	
		As a result, $ R1 $ discarded the MaxAge LSA, and re-sent to $ R2 $ the false LSA from its LSDB.
		Then, $ R1 $ kept sending to $ R2 $ the false LSA and $ R2 $ kept sending to $ R1 $ the MaxAge LSA.	

		\textbf{Impact:} This deviation allows an attacker to send a spoofed LSA that disrupts the routing in the network. During the attack a very long period of instability is observed, on which two routers keep exchanging repeated instances of LSAs. During that period the routers remain with inconsistent routing tables, and there is loss of connectivity between some of the routers.
		

\end{enumerate}

The above deviations manifested themselves even further when tests included two LSAs. We partially applied an analysis of depth two, for two new reachable states, on IOS versions 15.1 and 15.2.

The new reachable states that we chose are:

\begin{enumerate}
	\item  The LSDBs of all routers contain a spoofed LSA of R0 with an empty list of links.
	\item The LSDB of R1 contains its own LSA with $ MaxSeqNum $, and the other routers' LSDBs are missing the LSA of R1.
\end{enumerate}

The generated tests contained two sent LSAs, where the first LSA  leads to one of the above chosen reachable states.
We did not find any new deviations in this analysis.
However, we did observe test failures that were related to deviation \#1 specified in the previous section.
For instance, the following scenario was observed:

\begin{itemize}
	\item Sent LSA \#1: an LSA with MaxSeqNum was sent on behalf of $ R $ to $ R $.
	\item Sent LSA \#2: an LSA with MaxSeqNum was sent on behalf of another router $ R' $ to $ R $.
	\item Final state: In Cisco's implementation, in the final state both LSAs of $ R $ and $ R' $ were initialized with InitialSeqNum on all routers' LSDBs.
\end{itemize}
 The expected final state from the model was that only the LSA of $ R' $ would be initialized, and the LSA of $ R $ would remain unchanged (i.e., with MaxSeqNum on R's LSDB and missing from the other routers' LSDBs, as described in deviation \#1). This scenario demonstrates that the second LSA unexpectedly affected the state of the routers w.r.t. a different LSA of another router ($ R $) as well. The result was that $ R $ completed its expected procedure only after the second LSA was sent. This new observed behavior can be described as a more complete view of deviation \#1 that was initially found in the 1-depth analysis. Increasing the depth of the analysis has the potential to reveal additional consequences of previously found deviation, as demonstrated in this case.

\subsubsection{Known vulnerabilities }

Below is the detailed list of additional vulnerabilities found in version 15.1 only. All of these vulnerabilities were patched by Cisco by version 15.2.
In the appendix we include more details of this list with specific descriptions of how each deviation was found.

\begin{enumerate}

	  \setcounter{enumi}{2}
	
	\item \textbf{Inconsistent LSA with $ LSID \neq AR $ poisons LSDBs and routing tables:}
		
		\textbf{Description}: A false inconsistent LSA with $ LSID \neq AR $ was sent to a router $ R $, where the LSID was equal to $ R $'s ID.
		The false LSA unexpectedly replaced the correct LSA of $ R $ on its own LSDB and on other routers' LSDBs as well. The routing tables of these routers were re-calculated based on that false LSA, and consequently no OSPF-derived route existed in their routing tables

		\textbf{Impact}: This deviation allows an attacker to send a spoofed LSA that persistently disrupts the routing in the network.

	\item \textbf{Inconsistent LSA with lower sequence number causes a fight-back:}

		\textbf{Description:} A false inconsistent LSA with $ LSID \neq AR $ was sent to a router $ R $,  with $ LSID=R $ and $ AR = R' $.
		 An unexpected fight-back LSA was originated by $ R' $, even though the sent LSA had lower sequence number than its own LSA. This is not in accordance with the OSPF specification, which states that a new LSA (a fight-back) should be sent in response to a self-originating false LSA only if that false LSA is newer than the current LSA.

		\textbf{Impact:}	The impact is similar to the one described in the previous deviation, since the final state in both scenarios is similar. This scenario describes an additional deviation with respect to the previous one, but it has no additional effect on the calculated routing tables.

	\item \textbf{Inconsistent fight-back response:}
	
		\textbf{Description:}   A false LSA with $ LSID \neq AR $ was sent to a router $ R $. One of the routers originated a fight-back response. Then, unexpectedly, its neighbor kept re-sending to that router the original false LSA, and the other router kept sending a new fight-back response with an incremented sequence number. This behavior was repeated for many such iterations until stabilization. 

		\textbf{Impact:} The observed behavior includes the behavior described in deviations 3 and 4, but it also affects the stability of the routers. It was observed on  specific routers within the topology on several specific combinations of values for the LSA fields.

	\item \textbf{Inconsistent fight-back response for MaxSeq-1}
		
		\textbf{Description:} A false LSA with $ LSID \neq AR $  and $ seq=MaxSeq-1  $ was sent to a router $ R $.  One of the routers originated a fight-back with $ seq=MaxSeq $ and $ age=MaxAge $. Consequently, its neighbor kept sending the original false LSA  over and over again, and it took many such iterations until stabilization.

		\textbf{Impact:} The observed behavior includes the behavior observed in deviation 3, but also affects the stability of the routers.
		
		
		\end{enumerate}

\subsubsection{Non-vulnerabilities}	
As already noted, not all deviations must be security vulnerabilities. A deviating functionality can enhance the security of the implementation as compared to the standard. This is the case in one of the deviations we uncovered which we call ``Re-flooding of LSA arriving from DR by unicast". Due to space constraints this deviation is detailed in the appendix.

\subsection{Additional affected routers} \label{sec:results2}	
To further demonstrate the generality of our tool, we also employed it to analyze the OSPF implementation of the Quagga Routing Suite -- a popular open source routing software~\cite{Quagga}.
The analysis of Quagga consisted of the 94 generated test files (see Table ~\ref{tbl:results}).

We found one significant deviation in this version that also poses a security vulnerability. The deviation is similar to deviation 2 from Section~\ref{unknown} that we found in Cisco's implementations. We confirmed the existence of the deviation in the source code. The deviation manifested itself in Quagga is all routers (in Cisco only some of the routers exhibited the deviation). This is due to the fact that once a Quagga router receives an LSA it immediately floods it to its neighbors and only then checks if it was originated by itself. This allows the false LSA to propagate to all routers before the fight-back LSA is triggered hence manifesting the described deviation in every router. This vulnerability affect all Linux flavors that package Quagga, e.g., SUSE, openSUSE and RadHat.

Subsequent to the disclosure of the deviations found in this work some of them were identified also by IBM~\cite{lenovo}, Lenovo~\cite{lenovo} and Huawei (known via private communication) in their own products.

\section{Advantages and limitations of our method} \label{advatagesnandlimitations}
Our method allows us to focus on specific protocol functionality on which an exhaustive testing can be applied for any implementation of the protocol, and black box in particular.
It can be used to test various implementations of different versions and vendors, using the same  model and the same set of tests, per  chosen network topology.
Since the test generation is systematic and exhaustive with full coverage, it is very effective in finding deviations of a protocol from its standard. The effectiveness is demonstrated by the large number of deviations that we found on an OSPF implementation with a single symbolic LSA and a standard initial state.

Furthermore, our tool can be easily adapted to search for deviation in a different vendor's implementation. One need  only adapt the commands required to fetch LSDB and routing table from the routers and parse them as needed. Adapting our method to a different network protocol is also straightforward; however, it requires a new reference model and adjustments of the black box testing script, including the comparison method.

In Section~\ref{sec:relatedwork} we present some previous work related to black box analysis. Many past works used  automatic model inference.
When comparing black-box implementations with the standard of the protocol as we did here, the automatic inference approach  would require applying such inferences for every new version that needs to be tested. Then, specific predefined properties would have to be tested on the inferred models.
In our approach a reference model  has to be implemented once, and its generated tests can be directly applied on every new implementation version, without inference.
It should also be noted that even though our method requires knowledge of the protocol and manual abstractions, such knowledge may also be required for the inference method.  The inference process may require non-trivial abstractions to enable inferring a certain part of the protocol in the form of a regular automaton with some abstracted alphabet. This abstraction process is non-trivial and requires knowledge of the protocol.


\paragraph{Limitations} The limitations of our approach include the manual implementation of a reference model. It is possible that the reference model has unintended deviations from the standard, therby undermining the credibility of the analysis.
The reference model should be compact, and thus must be limited to certain chosen functionalities of the protocol. The method may  have scalability issues for larger models. In such cases  heuristics and priorities may have to be developed for partial generation of test cases. Another option is to split a large model into several smaller compact models.

Furthermore, as for all black box analysis methods, our method cannot guarantee full coverage of the \emph{implementation code}. The generated tests only provide full coverage of the model.  Therefore, our method is not geared towards finding software vulnerabilities that stem from buffer overflows, race conditions, parsing errors and the like.

Additionally, despite our optimizations, the number of symbolic variables and their domains may grow along with any increase in the number of sent messages or the topology size, leading to the path explosion problem. In this paper, we do not explore the topology size limit or message depth limit of our method. Nonetheless, we have shown in previous section that large topologies are not necessarily required to expose non-trivial deviations.


\vspace{-0.2cm}

\section{Related Work} \label{sec:relatedwork}

\subsection{Fuzzing}
Fuzzing \cite{mcnally2012fuzzing} is a common black box testing approach. It is based on generation of random or faulty unexpected input. Such inputs are tested against the SUT to reveal implementation errors.

Some previous works have developed network protocol fuzzers for vulnerability analysis. SPIKE \cite{aitel2002introduction} is a framework for creating block based network fuzzers. Block-based fuzzing involves  splitting messages into static and dynamic parts, where fuzzing is applied on the dynamic parts.
In \cite{banks2006snooze}, a stateful fuzzing approach was implemented for the SIP protocol.
In \cite{schneider2013online},  a fuzzing approach that generated test cases at runtime was presented.  The authors generated behavioral fuzz test cases from UML sequence diagrams by applying a set of fuzzing operators.


Unlike random fuzzers that generally do not guarantee completeness of the analysis, our approach is designed to apply an exhaustive and systematic analysis for 
certain modeled functionalities.
Thus, our approach 
thoroughly tests a predefined set of protocol functionalities.


\subsection{Formal Black Box Analysis}

Some previous works have used model based approaches for black box analysis. For example, in \cite{zhuang2014netcheck}, a black box  analysis was applied on networked applications for fault detection
by analyzing traces of system calls.
The analysis was 
used to find deviations from expected network semantics on certain points in the ordered execution.
In \cite{bishop2005rigorous} a technique for rigorous protocol specification was developed and applied on the TCP and UDP protocols. The specification is written as operational semantics definition in higher order logic.
A specification-based testing approach is used to test some implementations. It is based on capturing SUT traces and using a checker, written above HOL, that performs symbolic evaluation of the rigorous specification along the captured traces.
In \cite{arcuri2010black} a black box approach was used by modeling real-time embedded systems environment in UML. 
The work focused on random testing, adaptive random testing and search-based testing.


\noindent
Other past works (\cite{fiteruau2014learning,hsu2008model,cho2010inference}) approached the black box analysis task using active learning algorithms (\cite{angluin1987learning,shahbaz2009inferring}) to automatically infer a model of the black box system in the form of an automaton.
In \cite{fiteruau2014learning} the inference algorithm was used to learn an automaton model for a fragment of the TCP protocol in two different implementations. The inferred models were then compared to obtain fingerprinting of these implementations. It should be noted that an abstraction of the TCP packets was used in the learning process. Thus, applying the method to a black box implementation still requires some knowledge of the protocol itself.
The authors of \cite {cho2010inference} used similar inference methods to learn models of botnet Command and Control protocols. To analyze the inferred model, they defined certain properties to check on the inferred state-machine.
In \cite{hsu2008model} a black box implementation of the MSN Instant Messaging Protocol is automatically inferred. A fuzz testing technique is used to analyze the inferred model and to search for inputs that can crash the implementation.

In \cite{argyros2016sfadiff} an approach based on automata learning is developed. It consists of a black-box differential testing framework based on Symbolic Finite Automata (SFA) learning. 
It is based on inferring  SFA models of the target programs using black-box queries and enumerating the differences between the inferred models. The method was evaluated on TCP implementations and Web Application Firewalls, and revealed differences between implementations.

\subsection{Symbolic Execution}

Symbolic execution is a very effective and common technique for test generation. It is used to analyze protocols (e.g. \cite{song2011rule},\cite{song2014symbexnet},\cite{anand2012automated}) mostly when the implementation is white box  with available source code. For example, in \cite{song2014symbexnet}, an analysis that uses symbolic execution is applied directly on the source code of network protocol implementations, such as DHCP and Zeroconf. It is used to test the protocol implementation against its specification. The specification from an RFC document is translated into a specification in a rule-based language. The input packets generated from the tests are then used to detect violations of the specified rules.







\subsection{OSPF analysis}

Several works have analyzed the OSPF standard itself for security vulnerabilities \cite{nakibly2014ospf,DBLP:conf/cav/SosnovichGN13, Jones06}. A few of these works have used the same threat model and even the same formal model of the OSPF standard, as we did here. However,  these works  used  a formal analysis process to identify security issues in the model: namely, they identify states in which the model has an undesirable property from a security point of view. In contrast, we use the formal model as a benchmark to test implementations, thereby allowing us to find security issues in specific implementations of the standard rather than in the standard itself.

A few earlier works have also addressed deviations in OSPF implementations. In ~\cite{Wang97, Wu99, vetter1997experimental}, the authors identify some deviations in OSPF implementations. All of the identified deviations related to incorrectly wrapping of the LS sequence number field,  potentially causing false LSAs to remain in the LSDB. The deviations found there are a result of an ad-hoc manual analysis.

\subsection{Deviations analysis}

There are some past works that suggested approaches for finding deviations between implementations of the same protocol. In ~\cite{brumley2007towards}, the suggested approach consists of automatically building symbolic formulas from the given implementation programs. Deviations between the implementations are identified by solving formulas generated from the two implementations. The approach was evaluated on the HTTP and NTP protocols.
Our approach, on the other hand, is focused at finding deviations between an implementation and a reference model. It does not require access to the implementation code or to its binary code. In addition, our approach focuses on complex multi-party routing protocols.

In ~\cite{brubaker2014using}, the authors developed a method for adversarial testing of certificate validation
logic in SSL/TLS implementations. It is based on frankencerts, synthetic certificates randomly mutated from parts of real certificates. They applied differential testing with frankencerts and found implementation flaws. 
The suggested method focuses on a specific protocol and the analysis was based on the new concept of frankencerts.

\section{Conclusions} \label{sec:conclusions}

In this work we developed and implemented a black-box method to find deviations of a closed-source protocol implementation from its standard. We used a model based approach in which we modeled core parts of the protocol's standard and used it as a reference model. We have shown that the method is efficient and practical for finding deviations in complex multi-party protocols. We did so by applying the method to the complex and widely used OSPF routing protocol. We tested three versions of Cisco's implementation and found different deviations in each. The method uses concolic execution to generate tests with high coverage, and thus it allowed us to find 7 significant deviations of the tested implementations even in relatively simple topologies. Most of these deviations 
pose security vulnerabilities and they were all confirmed by Cisco.
By applying our method to the OSPF implementation of the Quagga Routing Suite and revealing a significant deviation, we further demonstrate the generality of the method.

\bibliographystyle{acm}
\bibliography{bibdb}

\begin{thebibliography}{10}

\bibitem{gns3}
Graphical network emulator v2.0.1.
\newblock http://www.gns3.net, 2016.

\bibitem{lenovo}
Industry-wide ospf routing vulnerability on lenovo and ibm networking switches.
\newblock \url{https://support.lenovo.com/il/en/product_security/len-14078},
  2017.

\bibitem{Quagga}
Quagga routing suite v1.2.1.
\newblock \url{http://www.nongnu.org/quagga/}, 2017.

\bibitem{scapy}
Scapy v2.3.2.
\newblock \url{http://www.secdev.org/projects/scapy/}, 2017.

\bibitem{adamczyk2008non}
{\sc Adamczyk, P., Hafiz, M., and Johnson, R.~E.}
\newblock Non-compliant and proud: A case study of http compliance.

\bibitem{aitel2002introduction}
{\sc Aitel, D.}
\newblock An introduction to spike, the fuzzer creation kit.
\newblock {\em presentation slides), Aug 1\/} (2002).

\bibitem{anand2012automated}
{\sc Anand, S., Naik, M., Harrold, M.~J., and Yang, H.}
\newblock Automated concolic testing of smartphone apps.
\newblock In {\em Proceedings of the ACM SIGSOFT 20th International Symposium
  on the Foundations of Software Engineering\/} (2012), ACM, p.~59.

\bibitem{angluin1987learning}
{\sc Angluin, D.}
\newblock Learning regular sets from queries and counterexamples.
\newblock {\em Information and computation 75}, 2 (1987), 87--106.

\bibitem{apfelbaum1997model}
{\sc Apfelbaum, L., and Doyle, J.}
\newblock Model based testing.
\newblock In {\em Software Quality Week Conference\/} (1997), pp.~296--300.

\bibitem{arcuri2010black}
{\sc Arcuri, A., Iqbal, M.~Z., and Briand, L.}
\newblock Black-box system testing of real-time embedded systems using random
  and search-based testing.
\newblock In {\em IFIP International Conference on Testing Software and
  Systems\/} (2010), Springer, pp.~95--110.

\bibitem{argyros2016sfadiff}
{\sc Argyros, G., Stais, I., Jana, S., Keromytis, A.~D., and Kiayias, A.}
\newblock Sfadiff: Automated evasion attacks and fingerprinting using black-box
  differential automata learning.
\newblock In {\em Proceedings of the 2016 ACM SIGSAC Conference on Computer and
  Communications Security\/} (2016), ACM, pp.~1690--1701.

\bibitem{banks2006snooze}
{\sc Banks, G., Cova, M., Felmetsger, V., Almeroth, K., Kemmerer, R., and
  Vigna, G.}
\newblock Snooze: toward a stateful network protocol fuzzer.
\newblock In {\em International Conference on Information Security\/} (2006),
  Springer, pp.~343--358.

\bibitem{bishop2005rigorous}
{\sc Bishop, S., Fairbairn, M., Norrish, M., Sewell, P., Smith, M., and
  Wansbrough, K.}
\newblock Rigorous specification and conformance testing techniques for network
  protocols, as applied to tcp, udp, and sockets.
\newblock In {\em ACM SIGCOMM Computer Communication Review\/} (2005), vol.~35,
  ACM, pp.~265--276.

\bibitem{brubaker2014using}
{\sc Brubaker, C., Jana, S., Ray, B., Khurshid, S., and Shmatikov, V.}
\newblock Using frankencerts for automated adversarial testing of certificate
  validation in ssl/tls implementations.
\newblock In {\em Security and Privacy (SP), 2014 IEEE Symposium on\/} (2014),
  IEEE, pp.~114--129.

\bibitem{brumley2007towards}
{\sc Brumley, D., Caballero, J., Liang, Z., Newsome, J., and Song, D.}
\newblock Towards automatic discovery of deviations in binary implementations
  with applications to error detection and fingerprint generation.
\newblock In {\em Usenix Security\/} (2007).

\bibitem{cadar2013symbolic}
{\sc Cadar, C., and Sen, K.}
\newblock Symbolic execution for software testing: three decades later.
\newblock {\em Communications of the ACM 56}, 2 (2013), 82--90.

\bibitem{chen2006can}
{\sc Chen, P.~Y., and Forman, C.}
\newblock Can vendors influence switching costs and compatibility in an
  environment with open standards?
\newblock {\em MIS Quarterly: Management Information Systems 30}, SPEC. ISS.
  (2006), 541--562.

\bibitem{cho2010inference}
{\sc Cho, C.~Y., Shin, E. C.~R., Song, D., et~al.}
\newblock Inference and analysis of formal models of botnet command and control
  protocols.
\newblock In {\em Proceedings of the 17th ACM conference on Computer and
  communications security\/} (2010), ACM, pp.~426--439.

\bibitem{z3}
{\sc De~Moura, L., and Bj{\o}rner, N.}
\newblock Z3: An efficient smt solver.
\newblock In {\em Proceedings of the Theory and Practice of Software, 14th
  International Conference on Tools and Algorithms for the Construction and
  Analysis of Systems\/} (2008), Springer-Verlag, pp.~337--340.

\bibitem{fiteruau2014learning}
{\sc Fiter{\u{a}}u-Bro{\c{s}}tean, P., Janssen, R., and Vaandrager, F.}
\newblock Learning fragments of the tcp network protocol.
\newblock In {\em International Workshop on Formal Methods for Industrial
  Critical Systems\/} (2014), Springer, pp.~78--93.

\bibitem{godefroid2005dart}
{\sc Godefroid, P., Klarlund, N., and Sen, K.}
\newblock Dart: directed automated random testing.
\newblock In {\em ACM Sigplan Notices\/} (2005), vol.~40, ACM, pp.~213--223.

\bibitem{hsu2008model}
{\sc Hsu, Y., Shu, G., and Lee, D.}
\newblock A model-based approach to security flaw detection of network protocol
  implementations.
\newblock In {\em Network Protocols, 2008. ICNP 2008. IEEE International
  Conference on\/} (2008), IEEE, pp.~114--123.

\bibitem{Jones06}
{\sc Jones, E., and Moigne, O.~L.}
\newblock {OSPF Security Vulnerabilities Analysis}.
\newblock Internet-Draft draft-ietf-rpsec-ospf-vuln-02, IETF, June 2006.

\bibitem{mcnally2012fuzzing}
{\sc McNally, R., Yiu, K., Grove, D., and Gerhardy, D.}
\newblock Fuzzing: the state of the art.
\newblock Tech. rep., DTIC Document, 2012.

\bibitem{ospf_rfc}
{\sc Moy, J.}
\newblock {OSPF} version 2.
\newblock IETF RFC 2328, Apr. 1998.

\bibitem{NDSS12}
{\sc Nakibly, G., Kirshon, A., Gonikman, D., and Boneh, D.}
\newblock Persistent {OSPF} attacks.
\newblock In {\em Proceedings of NDSS\/} (2012).

\bibitem{nakibly2014ospf}
{\sc Nakibly, G., Sosnovich, A., Menahem, E., Waizel, A., and Elovici, Y.}
\newblock Ospf vulnerability to persistent poisoning attacks: A systematic
  analysis.
\newblock In {\em Proceedings of the 30th Annual Computer Security Applications
  Conference\/} (2014), ACM, pp.~336--345.

\bibitem{virl}
{\sc Obstfeld, J., Knight, S., Kern, E., Wang, Q.~S., Bryan, T., and Bourque,
  D.}
\newblock Virl: the virtual internet routing lab.
\newblock In {\em ACM SIGCOMM Computer Communication Review\/} (2014), vol.~44,
  pp.~577--578.

\bibitem{schneider2013online}
{\sc Schneider, M., Gro{\ss}mann, J., Schieferdecker, I., and Pietschker, A.}
\newblock Online model-based behavioral fuzzing.
\newblock In {\em Software Testing, Verification and Validation Workshops
  (ICSTW), 2013 IEEE Sixth International Conference on\/} (2013), IEEE,
  pp.~469--475.

\bibitem{sen2006cute}
{\sc Sen, K., and Agha, G.}
\newblock Cute and jcute: Concolic unit testing and explicit path
  model-checking tools.
\newblock In {\em International Conference on Computer Aided Verification\/}
  (2006), Springer, pp.~419--423.

\bibitem{shahbaz2009inferring}
{\sc Shahbaz, M., and Groz, R.}
\newblock Inferring mealy machines.
\newblock In {\em International Symposium on Formal Methods\/} (2009),
  Springer, pp.~207--222.

\bibitem{song2014symbexnet}
{\sc Song, J., Cadar, C., and Pietzuch, P.}
\newblock Symbexnet: testing network protocol implementations with symbolic
  execution and rule-based specifications.
\newblock {\em IEEE Transactions on Software Engineering 40}, 7 (2014),
  695--709.

\bibitem{song2011rule}
{\sc Song, J., Ma, T., Cadar, C., and Pietzuch, P.}
\newblock Rule-based verification of network protocol implementations using
  symbolic execution.
\newblock In {\em Computer Communications and Networks (ICCCN), 2011
  Proceedings of 20th International Conference on\/} (2011), IEEE, pp.~1--8.

\bibitem{DBLP:conf/cav/SosnovichGN13}
{\sc Sosnovich, A., Grumberg, O., and Nakibly, G.}
\newblock Finding security vulnerabilities in a network protocol using
  parameterized systems.
\newblock In {\em Computer Aided Verification - 25th International Conference,
  {CAV} 2013, Saint Petersburg, Russia, July 13-19, 2013. Proceedings\/}
  (2013), pp.~724--739.

\bibitem{utting2012taxonomy}
{\sc Utting, M., Pretschner, A., and Legeard, B.}
\newblock A taxonomy of model-based testing approaches.
\newblock {\em Software Testing, Verification and Reliability 22}, 5 (2012),
  297--312.

\bibitem{vetter1997experimental}
{\sc Vetter, B., Wang, F., and Wu, S.~F.}
\newblock An experimental study of insider attacks for ospf routing protocol.
\newblock In {\em Network Protocols, 1997. Proceedings., 1997 International
  Conference on\/} (1997), IEEE, pp.~293--300.

\bibitem{Wang97}
{\sc Wang, F., Vetter, B., and Wu, S.~F.}
\newblock {Secure Routing Protocols: Theory and Practice}.
\newblock Tech. rep., North Carolina State University, May 1997.

\bibitem{mini_mc}
{\sc Wang, X.}
\newblock mini-mc.
\newblock \url{https://github.com/xiw/mini-mc}, 2015.

\bibitem{Wu99}
{\sc Wu, S.~F., and et. al.}
\newblock Jinao: Design and implementation of a scalable intrusion detection
  system for the ospf routing protocol.
\newblock {\em ACM Transactiom on Computer Systems, VoL 2\/} (1999), 251--273.

\bibitem{zhuang2014netcheck}
{\sc Zhuang, Y., Gessiou, E., Portzer, S., Fund, F., Muhammad, M.,
  Beschastnikh, I., and Cappos, J.}
\newblock Netcheck: Network diagnoses from blackbox traces.
\newblock In {\em 11th USENIX Symposium on Networked Systems Design and
  Implementation (NSDI 14)\/} (2014), pp.~115--128.

\end{thebibliography}

\appendix



	

\section{Detailed description of deviations} \label{sec:results_detailed}

Table~\ref{tbl:results_with_topology_2} summarizes the number of generated tests per topology and model version.

In Topology 2 we tested the implementation of versions 15.2 and 15.6, and no new deviations were found. Two of the found deviations (1,2) were reproduced in this topology.

\begin{table*}
	\centering
	\begin{tabular}{|c|c|c|c|}
		\hline Topology & IOS veresion & \# Generated tests & \# Found deviations  \\
		\hline  Topology 1 & version 15.1 & 395  &  7 \\
		\hline  Topology 1 & version 15.2,15.6 & 395  & 3 \\ 	
		\hline Topology 2 & version 15.2,15.6 & 395 & 2 \\
		\hline
	\end{tabular}
	\caption{A summary of the number of generated tests and found deviations per each topology and IOS version with a single symbolic LSA}
	\label{tbl:results_with_topology_2}
\end{table*}

In this section we give more specific details on each of the found deviations.

\subsubsection{Results on Topology 1 with version 15.1}

We started with Topology 1 on version 15.1. The model was configured with one symbolic LSA. We found the following deviations as detailed below.
\begin{enumerate}
	\item \textbf{Incomplete fight-back for Rogue LSA with maximum sequence number:}
	
	\begin{itemize}
		\item \textbf{Sent LSA:} $ \langle src=R0,dest=R1,LSID=R1,AR=R1,seq=MaxSeq, links=[]  \rangle $
		\item\textbf{ Description:} A false LSA having the maximum sequence number was sent from Cloud 1 to $ R1 $. The false LSA had an LSID and Advertising Router fields that correspond to $ R1 $ ID. In response R1 flooded over all its interfaces a new LSA having maximum sequence number and maximum age in order to flush the false LSA from all the routers in the network. This action is expected and is in accordance with the OSPF specification. Following the reception of the LSA Acks from all its neighbors R1 was expected to send yet another LSA, but this time with the minimum sequence number. However, R1 does not send this LSA. As a result, the LSA of R1 is not installed in the LSDB of the routers in the network, causing their LSDB to be inconsistent with the actual network topology.
		\item \textbf{Impact:}  This gap allows an attacker to send a spoofed LSA that persistently harms the routing in the network.
		\item \textbf{Comments:} The same behavior was observed on the corresponding tests of $ R0 $ and $R2 $ (with $ dest=LSID=AR=R0 $ or $ R2 $), but on routers $ R3$ and $R4 $ the expected behavior was observed on this topology.
		\item \textbf{Status:} This deviation was acknowledged by Cisco.
	\end{itemize}

		\item \textbf{Incorrect MaxAge LSA origination during fight-back:}
		\begin{itemize}
			\item \textbf{Sent LSA:}  $ \langle src=R0,dest=R1,LSID=R2,AR=R2,seq=MaxSeq, links=[]  \rangle $
			\item \textbf{Description:}  A false LSA having the maximum sequence number was sent on behalf of $ R2 $ to $ R1 $. Consequently, $ R1 $ stored the false LSA in its LSDB and flooded it to $ R2 $, as expected. Since $ R2 $ received a self-originated LSA with MaxSeqNum, it should first originate a MaxAge LSA, and then an InitialSeqNum LSA.
			Based on the RFC, the MaxAge LSA should carry the same links as of the MaxSeqNum LSA.
			However, in the tested implementations, $ R2 $ originated a MaxAge LSA with its own valid links, instead of the invalid links from the false LSA.
			After $ R2 $ sent this wrong MaxAge LSA to $ R1 $, $ R1 $ had to check whether the MaxAge LSA is newer than the false LSA in its LSDB or not. Both LSAs have the same sequence number, which is MaxSeq.
			Based on the RFC, the checksum field is compared in such case. The false LSA in $ R1 $'s LSDB had a larger checksum value in our test. Thus, this false LSA was considered newer.
			Note that if the MaxAge LSA would have been correct, the checksum values would have been identical, and the MaxAge LSA would have been considered newer, as expected.
			
			As a result, $ R1 $ discarded the MaxAge LSA, and re-sent to $ R2 $ the false LSA from its LSDB.		
			
			Then, $ R1 $ kept sending to $ R2 $ the false LSA and $ R2 $ kept sending to $ R1 $ the MaxAge LSA.	
			In addition, the following message showed up in the console: ``Detected router with duplicate router ID'' (with RID of $ R2 $).

			\item \textbf{Impact:} This deviation allows an attacker to send a spoofed LSA that harms the routing in the network. During the attack a very long period of instability is observed, on which two routers keep exchanging repeated instances of LSAs. During that period the routers remain with inconsistent routing tables, and there is loss of connectivity between some the routers.
			
			\item \textbf{Comments:} The same behavior is observed when a similar LSA is sent with any $ dest $ that does not equal $ R2 $'s ID.
			\item \textbf{Status:} This deviation was acknowledged by Cisco.
		\end{itemize}	
	
	\item \textbf{Inconsistent LSA with $ LSID \neq AR $ poisons LSDBs and routing tables:}
	\begin{itemize}
		\item \textbf{Sent LSA:} $ \langle src=R0,dest=R1,LSID=R1,AR=R4,seq=n , links=[]  \rangle $
		
		\item \textbf{Description}: The false LSA sent to $ R1 $  had an LSID with $ R1 $'s ID and Advertising Router with $ R4 $'s ID. In the initial state the sequence number of $ R4 $'s LSA is less than $ n $.
		In the final state the LSA originated by $ R1 $ is unexpectedly replaced with the sent LSA at the LSDBs of  $R1,R2,R3  $. Thus, these LSDBs remain poisoned at the end of this test. This is not in accordance with the OSPF specification which says that an LSA is identified by the fields $ LSID,AR $, and type. The sent LSA should not have replaced the LSA of R1 since their $ AR $ values are different.
		$ R4 $ responds with a fight-back since the sent LSA contains $ AR=R4 $. Thus, its LSA's sequence number is increased and its value is $ n+1 $ on the final state for all routers' LSDBs.

		\item \textbf{Impact}: Following this attack the routing tables of R3 and R2 are re-calculated according to the false LSA and consequently no OSPF-derived route exists in their routing tables. The routing tables of R1 and R4 are unaffected by the attack.
		
		\item \textbf{Status:} This gap was already known and was published in \cite{nakibly2014ospf}. It was fixed on later versions. On the applied fix a router ignores  inconsistent LSA with $ LSID \neq AR $.

	\end{itemize}

	\item \textbf{Inconsistent LSA with lower sequence number causes a ‘fight-back’:}
	
	\begin{itemize}
		
		\item \textbf{Sent LSA:}  $ \langle src=R0,dest=R1,LSID=R1,AR=R2,seq=n , links=[]  \rangle $
		\item \textbf{Description:} A false LSA having $ LSID=R1 $ and $ AR=R2 $ was sent to R1 from Cloud 1. This LSA has a sequence number $ n $ that was larger than the seq of $ R1$’s initial LSA, but smaller than the seq of $ R2 $’s initial LSA. R1 floods the false LSA to all its neighbors including R2.
		R2 initially sends a fight-back LSA with a seq smaller than the seq of its own LSA currently installed in its DB (it simply sent an LSA that has a seq  increased by 1 compared to the seq of the false LSA). This is not in accordance with the OSPF specification which says that a new LSA (a fight-back) should be sent in response to a self-originating false LSA only if that false LSA is newer than the current LSA (see Sec. 13.1 in the RFC). This is not the case in our test.  As noted, the false LSA had a sequence number that is smaller than that of the current LSA.
		Eventually, R2 originates a new fight-back LSA with a seq that is increased by 1 compared to the LSA installed in its DB only after R1 sends to R2 the LSA with the updated seq.
		
		\item \textbf{Impact:}	The impact is similar to the one described in the previous deviation, since the final state on both scenarios is similar. This scenario describes an additional deviation with respect to the previous one, but it has no additional effect on the calculated routing tables.	
		\item \textbf{Status:}	We are not aware of any report about this deviation on version 15.1. However, due to the fix mentioned in the previous deviation, we could not reconstruct this deviation on later versions as well.	
		
	\end{itemize}

	\item \textbf{Inconsistent fight-back response:}
	\begin{itemize}
		\item \textbf{Sent LSA:} $ \langle src=R1,dest=R3,LSID=R0,AR=R4,seq=0x7, links=[]  \rangle $
		
		\item \textbf{Description:}   A false LSA having LSID=$ R0 $ and Advertising Router=$ R4 $ was sent to R3 from Cloud 1. It is sent with seq = 0x7. The false LSA is then flooded from $ R3 $ to $ R4 $. $ R4 $ replies with a fight-back having seq=0x8. In response, R3 sends to R4 the LSA with the updated seq=0x9, which is larger than the seq of the fightback (it was the initial sequence num of $ R4 $). Then, R4 sends an updated LSA with seq=0xA. Until this point the behavior is as previously described on gap 3 (R4 should not have sent a fight-back since the LSA sequence number was less than its own LSA ).
		Then, $ R3 $ unexpectedly re-sends to R4 the original (false) LSA with seq=0x7. Eventually, R4 sends an updated LSA with seq=0xB. This behavior goes on repeatedly, and for each such iteration R4 eventually sends an LSA with seq increased by 1. The last packet sent by R4 has a seq=0x16, and then there is a stable state.
		Additionally, the following message shows up:
		``Detected router with duplicate router ID'' with the ID of $ R4 $.

		\item \textbf{Impact:} The observed behavior includes the behavior observed on gap 3, but also affects the stability of the routers.
		
		\item \textbf{Comments:} The gap was only observed on several specific tests from all tests for which the failure was related to gap 3.
		
		\item \textbf{Status:} Since gap 3 was fixed on later versions, this gap was not observed on later versions as well.
	\end{itemize}
	\item \textbf{Inconsistent fight-back response for MaxSeq-1}
	\begin{itemize}
		\item \textbf{Sent LSA:}  $ \langle src=R1,dest=R4,LSID=R0,AR=R4,seq=MaxSeq-1, links=[]  \rangle $
		
		\item \textbf{Description:} We send a false LSA having LSID=R0 and Advertising Router=R4 to R4. The LSA had a seq=MaxSeq-1. R4 responds with a fight-back having seq=MaxSeq and age=MaxAge. However, R3 keeps sending the false LSA having seq=MaxSeq-1 over and over again, and it takes many iterations till stable state is achieved.
		Additionally, the following message shows up:
		``Detected router with duplicate router ID'' with ID of $ R4 $.

		\item \textbf{Impact:} The observed behavior includes the behavior observed on gap 2, but also affects the stability of the routers.
		
		\item \textbf{Comments:} The gap was only observed on several specific tests from all tests that match to gap 2, with seqNum = maxSeq-1.
		
		\item \textbf{Status:} Since gap 2 was fixed on later versions, this gap was not observed on later versions as well.
		
	\end{itemize}

	\item \textbf{Re-flooding of LSA arriving from DR by unicast:}
	
	\begin{itemize}
		\item \textbf{Sent LSA:}  $ \langle src=R1,dest=R3,LSID=R1,AR=R1,seq=n, links=[]  \rangle $
		
		\item \textbf{Description:} The above LSA is sent to $ R3 $. $ R3 $ receives it by unicast from $ R1 $. On the observed behavior, $ R3 $ floods the LSA by unicast to $ R4 $, and then $ R4 $ floods the LSA by unicast to $ R1 $. This results in a fight-back LSA that $ R1 $ originates. On the modeled behavior, $ R3 $ does not flood the sent LSA since it was sent from $ R1 $ which is the Designated Router. The RFC mentions the following: ``If the new LSA was received on this interface, and it was	received from either the Designated Router or the Backup Designated Router, chances are that all the neighbors have
		received the LSA already.  Therefore, examine the next 	interface.''
		Thus, our model follows the RFC instructions. However, based on the RFC, the flooding from the DR within a broadcast network is always expected to be by broadcast and not unicast: ``The only packets not sent as unicasts are on broadcast networks; on these networks Hello packets are sent to the multicast destination AllSPFRouters, the
		Designated Router and its Backup send both Link State Update Packets and Link State Acknowledgment Packets to the multicast address AllSPFRouters, while all
		other routers send both their Link State Update and Link State Acknowledgment Packets to the multicast address AllDRouters.''. The RFC assumes this is always true and does not contain any instruction to verify that.
		Thus, we infer that Cisco's implementation verifies whether the LSA packet sent from the DR was indeed flooded by multicast and not unicast as expected. If it is sent by unicast, they add the additional re-flooding as described on the observed behavior, to make sure that neighbors receive the LSA.

		\item \textbf{Impact:} This gap demonstrates improved security of the implementation with respect to the model which is based on the RFC. The final state on our model results in a poisoned LSDB of $ R3 $ with a fake LSA by $ R1 $, whereas on Cisco's implementation it is prevented by a fightback due to the re-flooding of the unicast LSA.

	\end{itemize}
	
\end{enumerate}

\subsubsection{Results on Topology 1 with versions 15.2, 15.6}

We re-applied the generated tests of topology 1 on the newer versions 15.2 and 15.6 of the OSPF Cisco's implementation.

On these versions we observed that deviations $ 3,4,5,6 $ from the previous section were not reproduced. That is due to the fix applied on this version, for which a router ignores inconsistent router LSA with $ LSID \neq AR $.
However, deviations $ 1,2,7 $ were still observed on this version.

\subsubsection{Results on Topology 2 with versions 15.2, 15.6}								
We used topology 2, which is a variation of topology 1. It required to update the model with the new topology, re-generate test files, and re-run the testing script with the new test files.
On this topology deviations $ 1,2 $ were still observed, and no other deviations were revealed.
As for deviation $ 1 $, on this topology the described  behavior was observed on all routers, and not only on $ R0,R1,R2 $ as on the previous topology.

\subsubsection{Results on depth 2 for Topology 1 with version 15.2}								

We partially applied our extension method on Topology 1 with version 15.2. Out of 94 tests of depth 1, only 9 result in a new reachable state. We applied a depth-2 sacn for two reachable states as  follows:
\begin{itemize}
	\item \textbf{The state:} The LSDB of all routers contains a spoofed LSA of R0 with an empy list of links. The corresponding LSA leading to that state is: $ [ dest=1, type= routerLSA, AdvertisingRouter= 0 , LSID= 0, sequenceNum= 1, Links= [] ]$
	
	\textbf{Results:} On this case we did not detect any new unexpected behavior. For every generated test with two LSAs where the first LSA is the above and the second results from the scan from the corresponding state, no gap on the expected beahvior was found.

	\item \textbf{The state:} The LSDB of R1 contained its own LSA with $ MaxSeqNum $, and the other routers LSDBs were missing R1's LSA. This state results from the gap described in the previous section. The corresponding LSA leading to that state is: $ [ src = 0:0 , dest=1, type= routerLSA, AdvertisingRouter= 1 , LSID= 1, sequenceNum= MAX, Links= [] ]$
	
	\textbf{Results:} On this case a gap was found: when the second LSA was sent on behalf of $ R4 $ with $ MaxSeqNum $ to $ R1 $, the final state contained both LSAs of $ R1 $ and $ R4 $ with $ initialSeqNum $ on all routers' LSDBs. The expected state from the model was that only the LSA of $ R4 $ would bt initialized, and the LSA of $ R1 $ would remain unchanged.
\end{itemize}

\end{document}